\documentclass[twocolumn]{aastex631}
\received{October 13, 2023}
\accepted{\today}

\begin{document}

\title{Predicting the Timing of the the Solar Cycle~25 Polar Field Reversal}

\author[0000-0003-3191-4625]{Bibhuti Kumar Jha}
\affiliation{Southwest Research Institute, Boulder, CO 80302, USA}

\author[0000-0003-0621-4803]{Lisa A. Upton}
\affiliation{Southwest Research Institute, Boulder, CO 80302, USA}

\correspondingauthor{Bibhuti Kumar Jha}
\email{maitraibibhu@gmail.com}



\begin{abstract}

The process of the Sun's polar field cancellation reversal commences with the emergence of new cycle Hale's polarity active regions. Once the Sun undergoes polarity reversal, typically occurring near the peak of solar activity, it begins the process of accumulating the seed field for the forthcoming solar cycle. In recent years, the Advective Flux Transport (AFT) model has proven highly effective in forecasting the progression of polar fields by leveraging observations of surface flows and magnetic flux emergence. In this study, we make use of the predictive capability of the AFT model to simulate the evolution of the polar fields and estimate the timing of the Solar Cycle 25 polarity reversal in both hemispheres of the Sun. We use the statistical properties of active regions along with Solar Cycle~13, which closely resembles the current solar cycle (Cycle~25), to generate synthetic active regions in order to simulate future magnetic flux emergence in AFT to predict the evolution of the polar field. Based on our simulations, we anticipate that the Northern hemisphere of the Sun will undergo a polarity reversal between June 2024 and November 2024, with the center of our distribution at August 2024. In the Southern hemisphere, we anticipate a polarity reversal between November 2024 and July 2025, centered around February 2025. Additionally, assuming that the reversal of the axial dipole moment coincides with the peak of the solar cycle, our findings indicate that Cycle 25 is expected to peak in 2024 (likely between April to August 2024).

\end{abstract}



\section{Introduction} \label{sec:intro}

Understanding solar activity cycle variability has been a persistent problem in the field of solar physics. Despite over a century of dedicated solar observations, this puzzle has yet to be resolved and our understanding of the solar activity cycle  remains incomplete \citep[see][]{Bhowmik2023}.
The intimate relation between the solar cycle and the polar field of the Sun was first suggested by \citet{Babcock1959} and put forward by \citet{Babcock1961} as the foundation of solar dynamo models \cite[see][for extensive review on the dynamo models]{Charbonneau2010}. The amplitude of the polar field at the beginning of the solar cycle acts as the seed field for the upcoming cycle and is one of the best proxies for predicting the strength of following solar cycle \citep{Svalgaard2005,Hathaway2010a, MunozJaramillo2012, Svalgaard2013, Upton2023a}. The reversal of the polar field occurs close to cycle maximum \citep{Babcock1959}, beginning the creation of the seed field for the upcoming cycle \citep{Golubeva2023}.

In the last few decades, Surface Flux Transport \citep[SFT;][]{DeVore1984, Wang1989, vanBallegooijen1998, Schrijver2001, Bhowmik2018} models have been exceptionally successful in simulating observed solar cycle behavior. SFT models illustrate how residual flux from tilted active regions (ARs) is carried to the poles by the meridional flow, leading to the cancellation of the existing polar field and the build up of the new polar field. However, the chaotic nature of the AR emergence and their tilts \citep{Jha2020} poses a challenge for models operating in a predictive mode. Without advance knowledge of future flux emergence, it is difficult to accurately predict the evolution of the polar field. 

A recent advancement in SFT modeling is the development of the Advective Flux Transport \citep[AFT;]{Upton2014, Upton2014a} model. AFT uses the observed flows on the Sun's surface, as opposed to parameterized flows. For example, diffusion is typically used in other SFT models to mimic the effects of convection \citep{Jiang2014a, Bhowmik2018, Yeates2023}, however AFT uses a convective simulation to explicitly incorporate the effects of the convective motions. AFT has proven successful in modeling the polar field evolution (e.g., obtaining an excellent match with the observed polar field from Wilcox Solar Observatory (WSO) and Helioseismic Magnetic Imager \citep[HMI;][]{Scherrer2012, Sun2015}, and has been reliable in predicting the timing of the Sun's polar field reversals \citep{Upton2014, Upton2014a, Hathaway2016}.

We are approaching the maxima of the current cycle (Solar Cycle~25) and the polarity reversal of the Sun's magnetic dipole is imminent. The evolution of the polar field in the near future is a marker for solar activity in the coming years, as we can expect the waning of solar activity after the reversal of polar fields. Knowing the level of solar activity in coming years is important for forecasting our space weather environment and ensuring the safety of our space technology and communication systems. In this letter we use the predictive capability of the AFT model to predict the timing of the polar field reversals in the Northern and Southern hemispheres of the Sun. We also estimate the phase lag in the timing of the reversals between the two hemispheres.

We briefly discuss the AFT model used for our predictions and provide an outline for how we use our knowledge of past solar cycles to create synthetic AR catalogs with the observed patterns of AR emergence in \autoref{sec:model}. In \autoref{sec:results} we presents our predictions of the timing of the hemispheric polarity reversal and the associated uncertainties based on different statistical techniques. Finally, in \autoref{sec:conclusion} we summarize our findings. 

\section{Surface Flux Transport Model}\label{sec:model}

AFT, like other SFT models, solves the radial component of the induction equation to  simulate the dynamics of the magnetic field on the surface of the Sun. The fundamental equation at the heart of the AFT is given by,
\begin{equation}
    \frac{\partial B_r}{\partial t} + \vec{\nabla} \cdot (\vec{u}B_r)=S(\theta, \phi, t)+\eta
\nabla^2B_r,
\label{eq1}
\end{equation}
 Here, $B_r$ is the radial component of magnetic field, and $u$ is the horizontal components of the surface flows, which includes axisymmetric flows (differential rotation and meridional flow) and convective flows \citep{Hathaway2011, RightmireUpton2012, Upton2014, Upton2014a}. The first term in the right hand side, $S(\theta, \phi, t)$, is the magnetic source term which represents new flux emergence at the solar surface. The second term, $\eta
\nabla^2B_r$ (where $\eta$ is diffusivity), is a diffusivity term added to stabilize the numerical scheme used in AFT and does not have any significant effect on the flux transport processes. See \citet{Upton2014, Upton2014a} for additional details about the model.

AFT can be operated in two different modes: baseline mode and predictive mode. In baseline mode, AFT uses data assimilation of magnetograms to produce the synchronic maps, representing an accurate snapshot of the Sun's entire photospheric magnetic field at a given time \citet{Upton2014, Upton2014a}. In predictive mode, AFT uses idealized bipolar ARs, to forecast the future evolution of the surface magnetic field. In the context of this letter, we create AFT Baseline maps by assimilating magnetograms from HMI up until 31st August, 2023. The Baseline map from 31st August, 2023 is then used as the initial condition to run the model further in time in the predictive mode. To run the AFT in predictive mode, we create ensembles of synthetic AR catalogs, based on the statistical properties of ARs and the timing and amplitude of previous solar cycles. ARs from these synthetic active catalogs are then incorporated into AFT as idealized bipolar magnetic ARs.

\subsection{Synthetic Active Regions Generator (SARG)}

To create synthetic AR catalogs, we use the Synthetic Active Regions Generator (SARG) code. To create a realization, SARG begins with the 13-month smoothed sunspot number v2.0 \citep{Clette2016}, taken from the Solar Influences Data Analysis Center (SIDC)\footnote{Monthly sunspot data is taken from \url{https://www.sidc.be/SILSO/home}.}. This is used to set the cadence of spot emergence. Here, SARG defines the number of days between subsequent AR emergence as $30.4368/(0.3 +  0.269736 \times SSN)$, where $SSN$ is the sunspot number v2.0 for a given month. For each AR, SARG draws on a random sample from the KPVT/SOLIS BMR Flux log-normal distribution ($\mu = 50.05~\&~\sigma = 0.75$ ) of flux as described in \citet{ MunozJaramillo2015,MunozJaramillo2021}. SARG randomly selects a hemisphere to place the spot and then determine the latitude of the AR by adding random fluctuations around the mean latitude location, which is given by the standard law for the equator-ward drift of the active latitudes as described in \citet{Hathaway2011a}. The longitude of the AR is then drawn from a random uniform distribution. SARG assigns the tilt of the AR based on the Gaussian distribution for Joy's Law detailed in \citep{Hale1919, MunozJaramillo2021}. The tilt and the separation distance \citep[in prep.]{Upton2023_aft304} determines the relative position of the bipoles for each AR. The polarity of each bipole is assigned based on the Hale's polarity law for that cycle and hemisphere \citep{Stenflo2012a}. Due to the inherent randomness in the observed properties, no two SARG realizations will yield the exact same set of ARs, even though their statistical properties are identical. For each SARG realizations, we incoporate the ARs into AFT as bipolar Gaussian spot pairs with the specified properties (date, flux, polarity, and location). 

\begin{figure}[hb]
    \centering
    \includegraphics[width=\columnwidth]{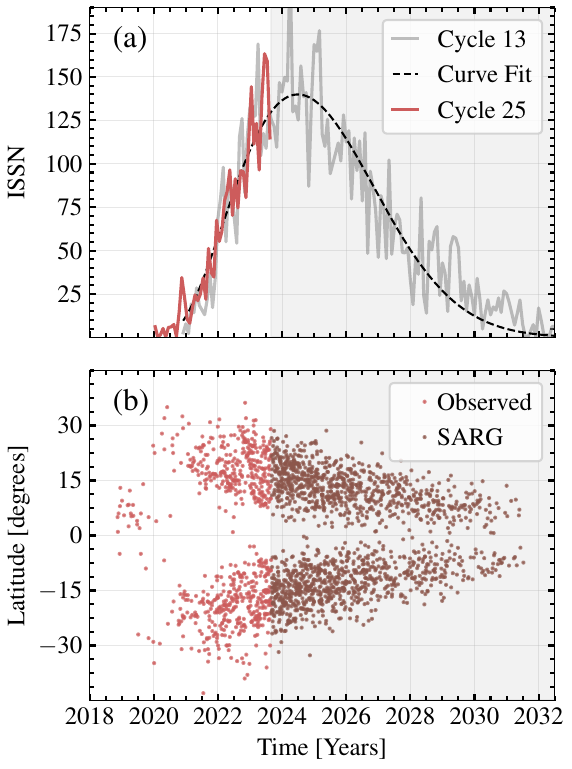}
    \caption{(a) The monthly average sunspot number for current cycle (Cycle~25), along with  monthly average sunspot number for Cycle~13 and the best fit curve based on \citet{Hathaway1994}. (b) shows the time latitude butterfly diagram for Cycle~25 up to August 2023 and one SARG realizations of synthetic ARs from September 2023 onward.}
    \label{fig1:cycle13}
\end{figure}

\subsection{Selection of Past Solar Cycle}

As discussed above, the selection ARs used for prediction is crucial. Here we base our SARG AR realizations on a past solar cycle that most closely resembles the current progress of Cycle~25 \citep[e.g., ] [ who used Cycle~14 in place of Cycle~24]{Hathaway2016}. In \autoref{fig1:cycle13}(a), we show the monthly averaged sunspot number v2.0 taken from the SIDC for Cycle~13 along with the current progress of Cycle~25. We fit an asymmetric curve \citep{Hathaway1994, Hathaway2011a, Upton2023} to the cycle and then shift it in time to match the timing of Cycle~25. As shown in \autoref{fig1:cycle13}(a), Cycle~13 is an excellent match for Cycle~25 in terms of monthly averaged sunspot number. The butterfly diagram shown in \autoref{fig1:cycle13}(b), further illustrates that the current cycle is closely following the SARG realization based on Cycle~13. This figure also illustrates that the frequency and distribution of ARs generated by SARG are qualitatively consistent with the observations. Hence, we select Cycle~13 as our reference cycle for SARG and produce 30 realizations of synthetic AR data. Simulating 30 different realizations in AFT highlights the potential variability due randomness inherent in the ARs. This allows us to characterize the uncertainty in our prediction of the polar field evolution.

\section{Results}\label{sec:results}

\begin{figure}[b]
    \centering
    \includegraphics[width=\columnwidth]{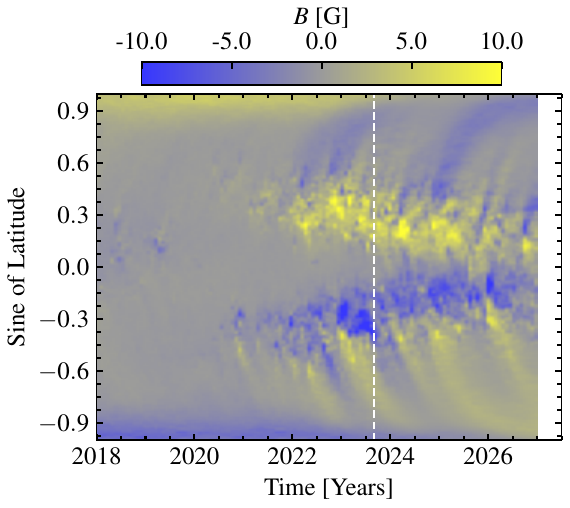}
    \caption{The magnetic butterfly diagram, constructed using AFT Baseline map till 31st August 2023 (marked using white dashed vertical line) and after that using one of the realizations of synthetic ARs in AFT's predictive mode.}
    \label{fig2:butterfly}
\end{figure}

Starting on September 1, 2023, we begin incorporating the SARG synthetic AR data into AFT and continue until the end of 2027. This process is repeated for all 30 SARG realizations. In \autoref{fig2:butterfly}, we show a magnetic butterfly diagram from one realization. The dashed white line indicates the transition of AFT from the baseline mode to the predictive mode. This figure shows how residual flux in ARs is the transported to the poles in streams of leading and following polarity flux. These streams drive the polar field evolution.

\begin{figure*}[t]
    \centering
    \includegraphics[width=\textwidth]{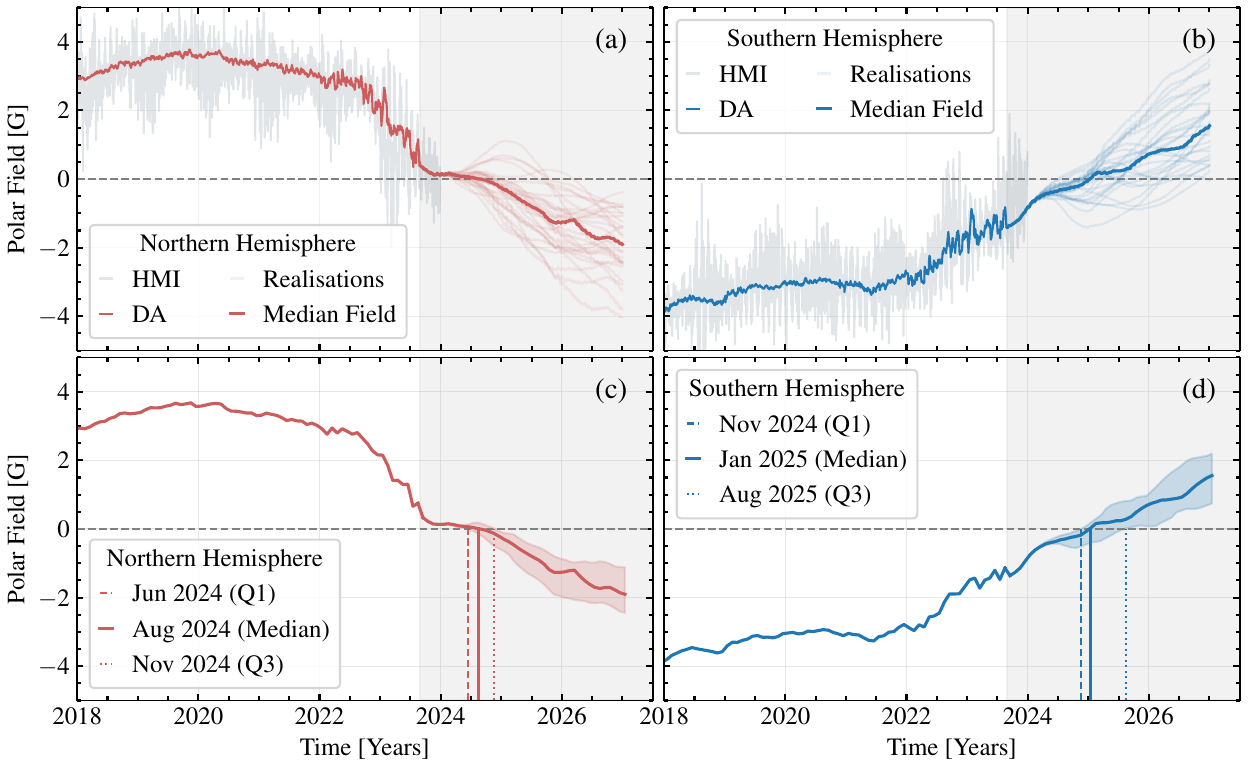}
    \caption{The polar field above $60\degr$ latitude for the (a) Northern and (b) Southern hemispheres is shown from the AFT Baseline through August 2023 (solid line), and for 30 SARG realizations afterward  (light color lines). For reference, the HMI polar field above $60\degr$ is also shown (light gray). The median of the 30 realizations is indicated by the solid line after August 2023. The median polar field measurements for the Northern (c) and Southern (d) hemisphere are shown in the same manner. Here, the shaded color regions represent the 50\% confidence interval between the first (Q1) and third (Q3) quartiles. The timing of the polar field reversal is marked by the vertical lines with the corresponding months noted in the legend.}
    \label{fig3:realizations}
\end{figure*}

The polar field is often calculated as the average magnetic flux density in the polar caps \citep{Upton2014}. WSO measures the polar field with a single pixel, nominally above $55\degr$ latitude\footnote{See \citet{Upton2023a} for a discussion on how the latitude range changes over the course of an orbit and the impact on the WSO polar field measurements.}, whereas HMI uses different latitude bands for this measurement, e.g. $50\degr-90\degr$ and $60\degr-90\degr$ \citep[see][for details]{Sun2015}. Here, we calculate the average polar fields above $60\degr$. In \autoref{fig3:realizations}a and \ref{fig3:realizations}b we show the polar field for the Northern and Southern hemispheres respectively. The polar field strength for all 30 realizations are shown as light color lines. The mean polar field for all realizations is indicted by the darker lines. For reference, we also include the HMI polar field measurements (light gray). We note that AFT shows excellent agreement with the HMI polar field.

We find no significant difference in the polar field of the 30 SARG realizations for nearly two years. This is expected as it typically takes a few years for the residual flux from the active latitudes to be transported to the poles. In \autoref{fig2:butterfly} we see that a negative polarity flux stream begins to migrate to the North pole around 6 months before we stop data assimilation This feature temporarily stalls the reversal of the Northern polar field. We note that this occurs in all of our simulations because the flux causing this unexpected behavior already exists on the Sun. Therefore, we can be confident that this will undoubtedly occur. While the Northern polar field stalls immediately, the Southern polar field initially continues its steady march toward reversal. However, we note that the ensemble of realizations do indicate that it may experience a brief stall of its own in 2024. This appears to be caused by a large concentration of negative flux in the active latitudes immediately before the assimilation process is stopped. While this will likely occur, it can be impacted by AR emergence in the coming months and is not as certain as the stalling of the Northern polar field. 

As we progress further in time, differences in the polar field evolution across realizations become more apparent and the polar field evolution of our simulations continues to diverge. This is confirmation that the chaotic nature of flux emergence makes the task of predicting polar field evolution during solar maximum for more that a few years into the future is a challenging task \citep{Golubeva2023}. However, as we near the polarity reversal, we can expect less uncertainty in the predictions. 

We now predict the of timing of polarity reversal in both the hemispheres by taking two different approaches, as discussed below.

\subsection{Uncertainty Based on the Median Polar Field}
We start by estimating the uncertainty in the timing of the reversal for the 30 different realizations used in this analysis. For each month, we compute the median polar field across all the realizations and calculate the first (Q1) and third (Q3) quartiles of the distribution. In \autoref{fig3:realizations}c and \ref{fig3:realizations}d, we show the temporal variation of this the polar field with 50\% confidence intervals( between Q1 and Q3, indicated by shaded color regions). We use the timing of the reversal of these curves (Q1, median and Q3) to get the expected time of polarity reversal and associated uncertainty. These reversal times are marked by vertical lines. This approach suggests that the Northern hemisphere is most likely to undergo a polarity reversal between June 2024 and November 2024, with the median time in August 2024. Conversely, the Southern hemisphere is expected to experience a polarity reversal between November 2024 and August 2025, with the median in January 2025.

\subsection{Uncertainty Based on the Individual Reversals}

\begin{figure}[htbp!]
    \centering
    \includegraphics[width=\columnwidth]{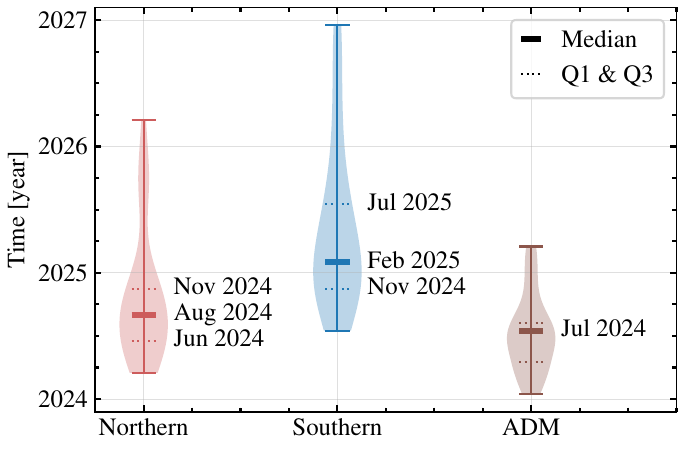}
    \caption{The violin plot shows the median timing of the polar field reversal, along with other statistical parameters such as the first (Q1) and third (Q3) quartiles, and the median timing of the reversal in the Northern (red) and Southern (blue) hemispheres. The third violin (brown) represents the distribution for the reversal of the axial dipole moment.}
    \label{fig4:quantiles}
\end{figure}

Next we discuss the second approach that we use to predict the timing of polarity reversal.
Here we calculate the timing of the polarity reversal for each individual realization in both hemispheres. We then use the distribution of these individual reversal to estimate the timing of the polarity reversal and the associated uncertainties. In \autoref{fig4:quantiles} we show the distribution of the timing for both hemispheres in the form of a violin plot, which is similar to the box plot \citep[see][for details]{Stryjewski2010}. This representation provides additional information about the probability distribution (shaded violin shaped region). Here, we use the Gaussian Kernel Density Estimator (KDE) to get the empirical probability distribution of the sample\footnote{See \url{https://matplotlib.org/stable/gallery/statistics/violinplot.html} for details about the violin plot.}. The extreme ends of the violins represent the range in the timing of polarity reversal based on our 30 realizations. Two dotted horizontal lines represent the first (Q1, 25th percentile) and third (Q3, 75th percentile) quartiles of the distribution. The solid lines represent the median of the distributions. After examining the distribution of timing, we find that the median is representative of the central tendency. Therefore, we use the median of the sample as our predictor and, Q1 and Q2 as our estimator of uncertainty for the timing of polarity reversal.

Based on the second approach, we predict that the Northern hemisphere is most likely to reverse it polarity in between June 2024 to November 2024 (50 percentiles), with median at August 2024. On the other hand, we predict that the Southern hemisphere will reverse its polarity sometime between November 2024 to July 2025 (50 percentiles), with median at February 2025. Using this apporach, we also evaluate the distribution for the timing of axial dipole moment \citep[ADM; see][]{Upton2014, Upton2014a} reversal. This indicates that the ADM is expected to change polarity in the middle of 2024. It's worth noting that the two distinct approaches used here exhibit good agreement in terms of the timing of polarity reversal, underlining that our predictions of polarity reversal timing is consistent and independent of the two methods.

As previously mentioned, different latitude limits can be used to calculate the polar fields. Therefore, we also calculate the timing of polarity reversals using our second approach with other latitude thresholds. In \autoref{tab:tab1}, we summarize the timing of polarity reversal for cases where lower latitude limits changes from $50\degr$ to $75\degr$ in increments of $5\degr$.

\begin{table*}[htbp!]
    \centering
    \begin{tabular}{|c|lll|lll|}
        \tableline
        & \multicolumn{3}{c|}{Northern Hemisphere}  & \multicolumn{3}{c|}{Southern Hemisphere} \\
         Latitude & Q1 & Median & Q3  & Q1 & Median & Q3 \\
         \tableline
         \tableline
        50$\degr$ & Jan 2024 & Apr 2024 & Jun 2024  & Nov 2023 & Feb 2024 & Dec 2024 \\ 
        55$\degr$ & Feb 2024 & Apr 2024 & Aug 2024  & May 2024 & Sep 2024 & Apr 2025 \\ 
        60$\degr$ & Jun 2024 & Aug 2024 & Nov 2024  & Nov 2024 & Feb 2025 & Jul 2025 \\ 
        65$\degr$ & Oct 2024 & Feb 2025 & Apr 2025  & Apr 2025 & Sep 2025 & Feb 2026 \\ 
        70$\degr$ & Feb 2025 & May 2025 & Aug 2025  & Aug 2025 & Dec 2025 & Jul 2026 \\ 
        75$\degr$ & May 2025 & Aug 2025 & Nov 2025  & Nov 2025 & Apr 2026 & Nov 2026 \\
         \tableline
    \end{tabular}
    \caption{The various statistical parameters for the timing of reversal of polar field for multiple latitude limits. Here, Q1 and Q3 represents the first and third quartiles of the distribution, respectively.}
    \label{tab:tab1}
\end{table*}

\section{Conclusion}\label{sec:conclusion}

Understanding the evolution of the polar field in the near future is important for gaining insights into solar activity. For example, the timing of this polarity reversal can provide an estimate of when to expect solar maximum. Once the sign of the Sun's polar field reverses polarity, the poles begin to build up magnetic flux of the opposite polarity, ultimately dictating the strength of the upcoming solar cycle. In this study we use the AFT model to predict the evolution of the polar field over the next few years. We simulate 30 realizations of synthetic ARs based on Solar Cycle 13 (which shows a good agreement with the current progress of Solar Cycle 25) as a proxy for the continued progression of the cycle. 
We use two different approaches to estimate the timing of polarity reversal in both hemispheres. Both approaches yield remarkably similar result in the prediction of the timing of the reversals. Consequently, we report the timing of polarity reversal based on the second approach, which uses the distribution of polarity reversal times across all 30 realizations. 

By measuring the average polar field above $60\degr$, we predict that for Cycle~25, the Northern hemisphere is likely to undergo a polarity reversal in August 2024 (with a 50\% confidence range spanning from June to November 2024). The Southern hemisphere is expected to reverse its polarity in February 2025 (with a 50\% confidence range from November 2024 to July 2025). Additionally, we conclude that for Cycle~25, the Northern hemisphere is expected to reverse its polarity $\approx 5 $ months before the Southern hemisphere, which is in line with the typical hemispheric lag. This is in stark contrast to Cycle~24, which was unusually asymmetric across the hemispheres and experienced a phase lag of approximately 16 months \citep{Sun2015}. Based on the assumption that the timing of the ADM reversal closely coincides with the time of solar cycle maximum, we also conclude that we are approaching the Solar Cycle 25 maximum and we can expect that solar activity will likely begin to decline in the second half of the 2024. This is consistent with the timing of solar maximum very recently reported in \citet{Upton2023a} based on the precursors method and current progress of the Cycle~25. However (\cite{Jaswal2024} suggest that the timing of the ADM reversal may not coincide with cycle maximum.

The findings of this study are important for advancing our capability of making solar cycle predictions. The approaches used in this work for predicting the evolution of polar field and quantifying the uncertainty associated with it, are important for accessing and determining our ability to use SFT models to make reliable predictions about the evolution of the polar field. Furthermore, they serve as a demonstration of our current understanding of the solar cycle and solar dynamo processes. Evaluating the precision and accuracy of these results after the polar field reversals have come to pass will be essential for determining how the stochastic nature of active region emergence limits our fundamental ability to make long term (many years) predictions.

\begin{acknowledgments}
BKJ and LAU were supported by NASA Heliophysics Living With a Star Strategic Capabilites grant NNH21ZDA001N-LWSS and NASA grant NNH18ZDA001N-DRIVE to the COFFIES DRIVE Center managed by Stanford University. LAU was also supported by NASA Heliophysics Living With a Star grant NNH18ZDA001N-LWS. HMI data used in this study are courtesy of NASA-SDO and the HMI science team. The authors would also like to thank Andrés Munoz-Jaramillo for his valuable suggestions and comments. This research has made use of NASA's Astrophysics Data System (ADS; \url{https://ui.adsabs.harvard. edu/}) Bibliographic Services. We also thank the anonymous referees whose useful suggestions improved the overall quality of the paper.
\end{acknowledgments}




% \appendix

% \section{Appendix information}


\software{Matplotlib \citep{Hunter2007}, Numpy \citep{harris2020array} and Pandas \citep{pandas2020}}

\bibliography{formated_all}{}
\bibliographystyle{aasjournal}



\end{document}